\documentclass[final,twoside,9pt]{IEEEtran}                 
\usepackage[utf8x]{inputenc}
\usepackage{graphicx}
\usepackage{amsmath}

\begin{document}

\title{A Multiscale Materials-to-Systems Modeling of Polycrystalline Pb-Salt Photodetectors}

\author{\IEEEauthorblockN{Samiran~Ganguly \IEEEauthorrefmark{1},~Moonhyung~Jang \IEEEauthorrefmark{1},~Yaohua~Tan \IEEEauthorrefmark{2},~Sung-Shik~Yoo \IEEEauthorrefmark{3},~Mool~C.~Gupta$^1$ \IEEEauthorrefmark{1},~Avik~W.~Ghosh \IEEEauthorrefmark{1},~\IEEEmembership{Senior Member,~IEEE}} 
\\ \IEEEauthorblockA{\IEEEauthorrefmark{1} Charles L. Brown Dept. of Electrical and Computer Engineering, University of Virginia, Charlottesville, VA 22904, USA\\ \IEEEauthorrefmark{2}Synopsys Inc., Mountain View, CA 94043, USA\\ \IEEEauthorrefmark{3}Northrop Grumman Corp., Rolling Meadows, IL 60008, USA}}

\maketitle

\begin{abstract} 
We present a physics based multiscale materials-to-systems model for polycrystalline $PbSe$ photodetectors that connects fundamental material properties to circuit level performance metrics. From experimentally observed film structures and electrical characterization, we first develop a bandstructure model that explains carrier-type inversion and large carrier lifetimes in sensitized films. The unique bandstructure of the photosensitive film causes separation of generated carriers with holes migrating to the inverted $PbSe|PbI_2$ interface, while electrons are trapped in the bulk of the film inter-grain regions. These flows together forms the 2-current theory of photoconduction that quantitatively captures the $I-V$ relationship in these films. To capture the effect of pixel scaling and minority carrier blocking, we develop a model for the metallic contacts with the detector films based on the relative workfunction differences. We also develop detailed models for various physical parameters such as mobility, lifetime, quantum efficiency, noise etc. that connect the detector performance metrics such as responsivity $\mathcal{R}$ and specific detectivity $\mathcal{D^*}$ intimately with material properties and operating conditions. A compact Verilog-A based SPICE model is developed which can be directly combined with advanced digital ROIC cell designs to simulate and optimize high performance FPAs which form a critical component in the rapidly expanding market of self-driven automotive, IoTs, security, and embedded applications.
\end{abstract}

\section{Introduction}

\IEEEPARstart{S}{olid}-state $Pb$-salt photodetectors, a century old technology \cite{rogalski_history_2012}, is undergoing a dramatic resurgence driven primarily through the rapidly emerging techno-economic trend of heterogenous integration. Erstwhile disparate fields of microelectronics like computation, memory, communication, and sensing are increasingly being integrated as monolithic systems \cite{soref_past_2006,bowers_silicon_2015,roelkens_mid-ir_2014,koppens_photodetectors_2014,roelkens_silicon-based_2013}. It is also becoming commonly accepted wisdom that the next generation of the ``Moore`s Law'' will see primacy of smart autonomous agents as the basic unit of technology\cite{waldrop_chips_2016}. These agents, be autonomous cars, self-contained medical devices, or increasingly powerful wireless handhelds and wearables, will almost always involve advanced smart sensors, including photon detectors sensing in various regions of electromagnetic spectrum, depending on the application. Due to close integration there will be performance demand on these devices not erstwhile sought. 

\begin{figure}
\center
  \includegraphics[width=3.3in]{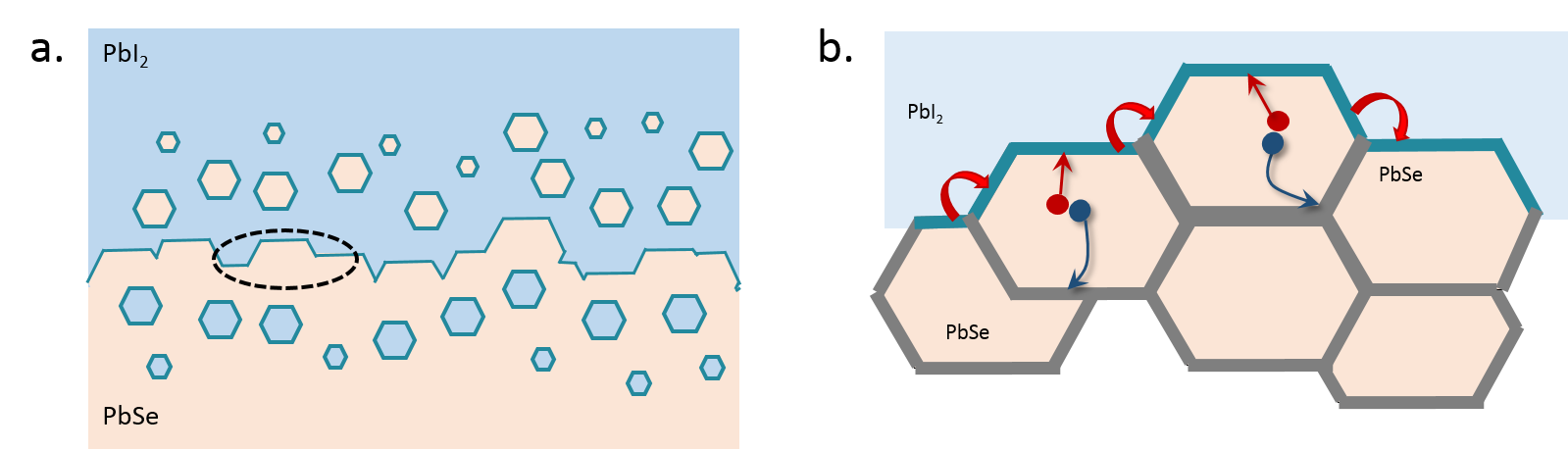}
\caption{a. Sensitized polycrystalline $PbSe$ film is composed to two stacked material layers: $PbSe$ and $PbI_2$ crystallites. The two materials form a rough interface due to diffusion, with individual isolated crystallites of the other material embedded in the bulk. b. Enlarged microscopic view of the indicated section. Wide band gap $PbI_2$ causes interfacial inversion of carrier type to which photo-generated holes are attracted, while electrons migrate to the bulk where they get trapped in the grain boundary regions. The holes conduct from source-to-drain by hopping through the islands of inverted interfaces.}
\label{fig:1}       
\end{figure}

These developments, therefore, call for a comprehensive physics based, materials-to-systems approach to modeling of photon sensors. In this work, we establish a framework for such an approach (fig.~\ref{fig:1}), with an illustrative and detailed model for $PbSe$ mid-wave infra-red (MWIR) photodetectors. While the model developed in this work is specific to this particular material system, the overall approach is general enough to be adapted for a wider class of photodetectors built using other material systems, due to its modularity.

The rest of the paper is organized as following. In section \ref{sec:material}, we describe the bandstructure of $PbSe$ films grown, sensitized, and characterized at Northrop Grumman Corp. and University of Virginia. We describe the carrier-type inversion and long carrier lifetimes in these films due to built-in large depletion fields formed during Iodization step in film fabrication. These films form a 2-layer heterostructure of $PbSe$ and $PbI_2$, giving rise to electrostatic doping of $PbSe$ due to band mismatch between the two materials. This forms the first part of the 2-current theory of photoconduction in these polycrystalline $PbSe$ films. We also discuss quantitative models of mobility and long carrier lifetime in these films arising from electrostatic doping. In section \ref{sec:transport}, we develop the second part of the 2-current theory using a Landauer-like model of hopping conduction of holes residing at the inverted interfacial islands of the $PbSe|PbI_2$ heterostructure. We also develop a Shockley-Quiesser based model for quantum efficiency that is quantitatively benchmarked with the experiments. In section \ref{sec:detector}, we develop a model for the noise in the detector and contact resistance based on the workfunction difference that incorporates the effect of current crowding seen in lateral contacts in ultrascaled transistors, as well as minority carrier blocking due to natural Schottky barriers formed at the film$|$contact interface. In section \ref{sec:circuit}, we describe a compact model for the detector that can be used in SPICE based circuit simulations and integrated with ROIC designs, illustrated using a few simulation examples. In section \ref{sec:projection}, we verify our complete model with two independent sets experiments by accurately benchmarking $I-V$, low bias $R-T$, and $\mathcal{D}^*-V$ using data provided by these experiments. We also make a few performance projections based on material improvements and operating conditions of these detectors. Finally we conclude in section \ref{sec:conclusion} with a few words on future directions for photosensor integration with advance neural filters.

\section{Detector Materials Model}
\label{sec:material}

During characterization on chemical bath deposition (CBD) grown and sensitized $PbSe$ films, we found \cite{yoo_high-operating-temperature_2017,jang_materials_2017,jang_notitle_nodate} that the bottom half of the film was MWIR sensitive polycrystalline $PbSe$ layer while the top half was converted to wide band-gap polycrystalline $PbI_2$ layer (fig. \ref{fig:1}). We also found that the $PbSe$ crystallites had a coating of nano-sized carbon particles which is likely to have come from the organic precursors during the deposition of the $PbSe$ film. 

Using Hall measurement studies, we also found that the conducting carriers in the as-grown as well as the oxidized films were from \textit{n}-type, whereas the Iodization caused carrier-type inversion, where the conducting carriers were of \textit{p}-type. This has been observed by others as well, for example see \cite{rogacheva_effect_2001,golubchenko_doping_2006}. However, we also found that stripping the top $PbI_2$ layer changed the conducting carriers to revert back to \textit{n}-type. Based on these observations, we develop the following model for the bandstructure of the sensitized CBD $PbSe$ material.

\subsection{Bandstructure, Electrostatic Doping, and Mobile Carrier-type Inversion}

\begin{figure}[t]
\center
  \includegraphics[width=3.3in]{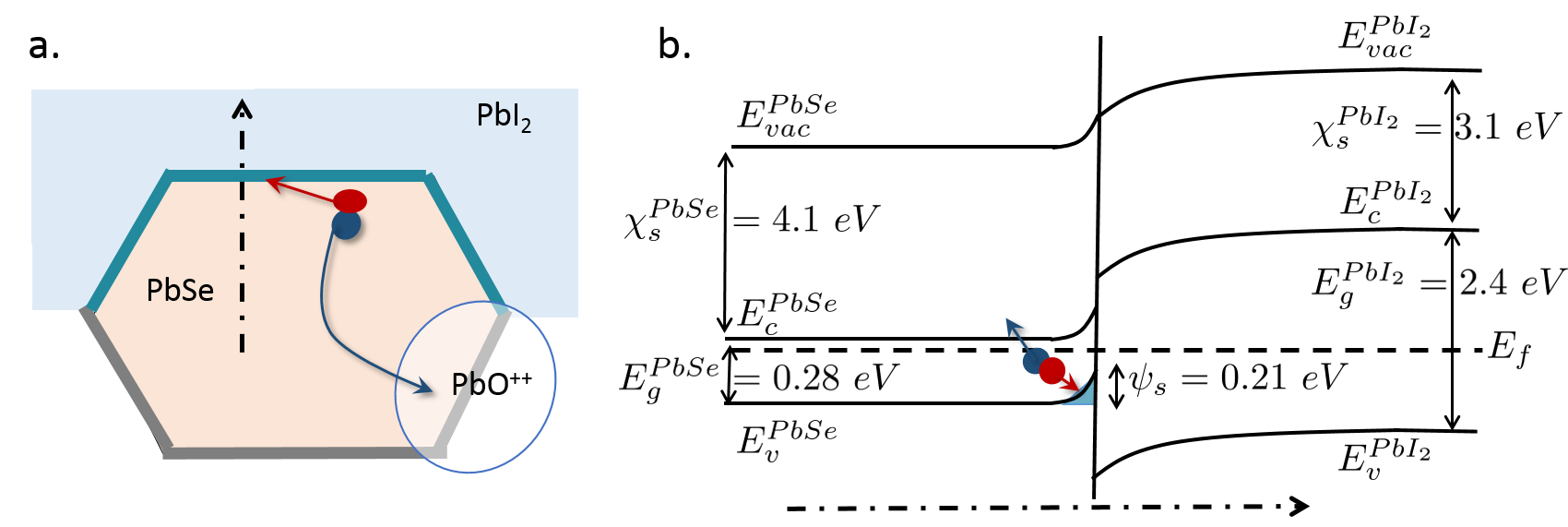}
\caption{a. Two layer photosensitive $PbSe|PbI_2$ heterostructure drawn for a single $PbSe$ crystallite. b. Heterostructure bandstructure (drawn along the dot-dash line) at the $PbSe|PbI_2$. Wide bandgap of intrinsic or \textit{p}-doped $PbI_2$ causes the \textit{interface of the $PbSe$ crystallite to be doped electrostatically to \textit{p}-type} forming a potential well (colored blue). During photo-excitation/carrier generation process, the depletion field separate the electron-hole pair and electrons migrate/diffuse to the bulk of the $PbSe$ and get trapped in the grain-boundary regions rich in vacancies ($PbO^{++}$)while the holes migrate to the inverted interface.}
\label{fig:2}       
\end{figure}

The central role of iodization in the sensitivity of the $PbSe$ detector and the reversible conducting carrier-type inversion points towards the importance of the $PbSe|PbI_2$ interface. While the polycrystalline materials do not have a well defined continuum bandstructure, we can nevertheless describe an individual crystallite (fig. \ref{fig:2}a) using a bulk single crystal bandsructure model, considering that the individual crystallites in this case are substantial in size ($d\sim 0.1-0.3\ \mu m$), and are close to the sizes of single crystal transistors which are described well with a bandstructure model. 

The heterostructure bandstructure of the $PbSe|PbI_2$ interface, drawn using the Anderson's Rule\cite{anderson_germanium-gallium_1960}, is shown in the fig.~\ref{fig:2}b, and supported by XPS data. We consider an intrinsic $PbI_2$ layer which is a much wider band gap material ($E_g \approx 2.4\ eV$) compared to $PbSe$ ($E_g \approx 0.28\ eV$)\cite{zemel_electrical_1965}. As a result, at the equilibrium due to Fermi level alignment, the normally \textit{n}-type $PbSe$ crystallite band bends towards a more \textit{p}-type region, giving rise to a depletion region and subsequently an inverted surface layer confined between the $PbI_2$ side band barrier and the bent bands of $PbSe$. 

The $PbSe|PbI_2$ interface is not sharp at the macroscopic level, due to inherent polycrystalline nature of the film and the band diagram drawn is at the individual crystallite level. Therefore, the inverted surfaces do not lie in the same geometrical plane but rather distributed along a band approximately equal to the characteristic diffusion length of Iodine into the $PbSe$ crystallites under the particular processing conditions. This can leave some isolated $PbSe$ crystallites embedded in the mostly  $PbI_2$ layer and isolated $PbI_2$ crystallites embedded in the mostly  $PbSe$ layer.

As an example of another work on crystallite surface depletion and corresponding potential profiles, where the depletion is assumed to cover the crystallites fully instead of being at the top interface, see \cite{bi_modeling_nodate}.

During electron-hole pair generation due to light incidence, this band alignment will cause the electrons to migrate towards the bulk of the $PbSe$ crystallite, while the holes move in to the inverted interface potential well (observed in many heterostructures)(fig.~\ref{fig:2}b). This inverted layer of holes is the primary charge carrier in the sensitized film, as observed during the Hall measurements. 

It should be noted that at the microscopic level, the degree of band-bending of each individual crystallite may vary depending on the detailed description of the crystallite doping, interfacial geometry etc. which can be captured by a finite element analysis for modeling purposes. However, for an analytical lumped parameter based estimate we use the measured doping in $PbSe$ ($2\times10^{15}\ cm^{-3}$) and doping in intrinsic $PbI_2$ ($1\times10^{11}\ cm^{-3}$) and estimate the depletion width in the $PbSe$ to be $ \simeq 17\ nm$, which is $\sim {1/10}^{th}$ the size the $PbSe$ crystallites ($d_{avg}\approx 200\ nm$), and a surface potential $\psi_s = 0.21\ eV$ at $T = 250\ K$, which is sufficient to support the carrier-type inversion ($PbSe$ band-gap $=0.28\ eV$). Since this band-bending and carrier inversion is caused by the electrostatics of the interface, it can be reversed by stripping the top $PbI_2$ layer which matches our observations as described before.

\subsection{Mobility}

For single crystal $PbSe$ films hole mobility has been found to be of a fairly high value of about $1000\ cm^2/Vs$ at room temperature \cite{allgaier_mobility_1958}, the polycrystalline films have typically much lower mobility due to heavy disorder and the measured mobility reduces to $0.1-1\ cm^2/Vs$ \cite{johnson_lead_1983}. From Hall measurements found that the hole mobility values  in the sensitized films were at an intermediate range of these two extremes to around $94\ cm^2/Vs$ \cite{jang_materials_2017,jang_notitle_nodate-1}, even though the film was highly polycrystalline. Additionally, it was found through HRTEM imaging that there is fairly large amount of carbon nano-particles present in the $PbSe$ grain boundaries as a result of $PbSe$ CBD and sensitization. 

It is well known that carbon compounds such as nanotubes and graphene have very high mobilities of $10,000\ cm^2/Vs$ and more at room temperature. We use a species fraction model of $PbSe$ and $C$ using a characterized by an ``effective mix ratio'' $\chi$ of the $C$ to obtain an effective mobility model given by:

\begin{equation}
\mu_{sp.fr.} = \chi\mu_C + (1-\chi)\mu_{PbSe}
\label{eq:mobility-sp-frac}
\end{equation}

\begin{figure}[t]
\center
  \includegraphics[width=3.3in]{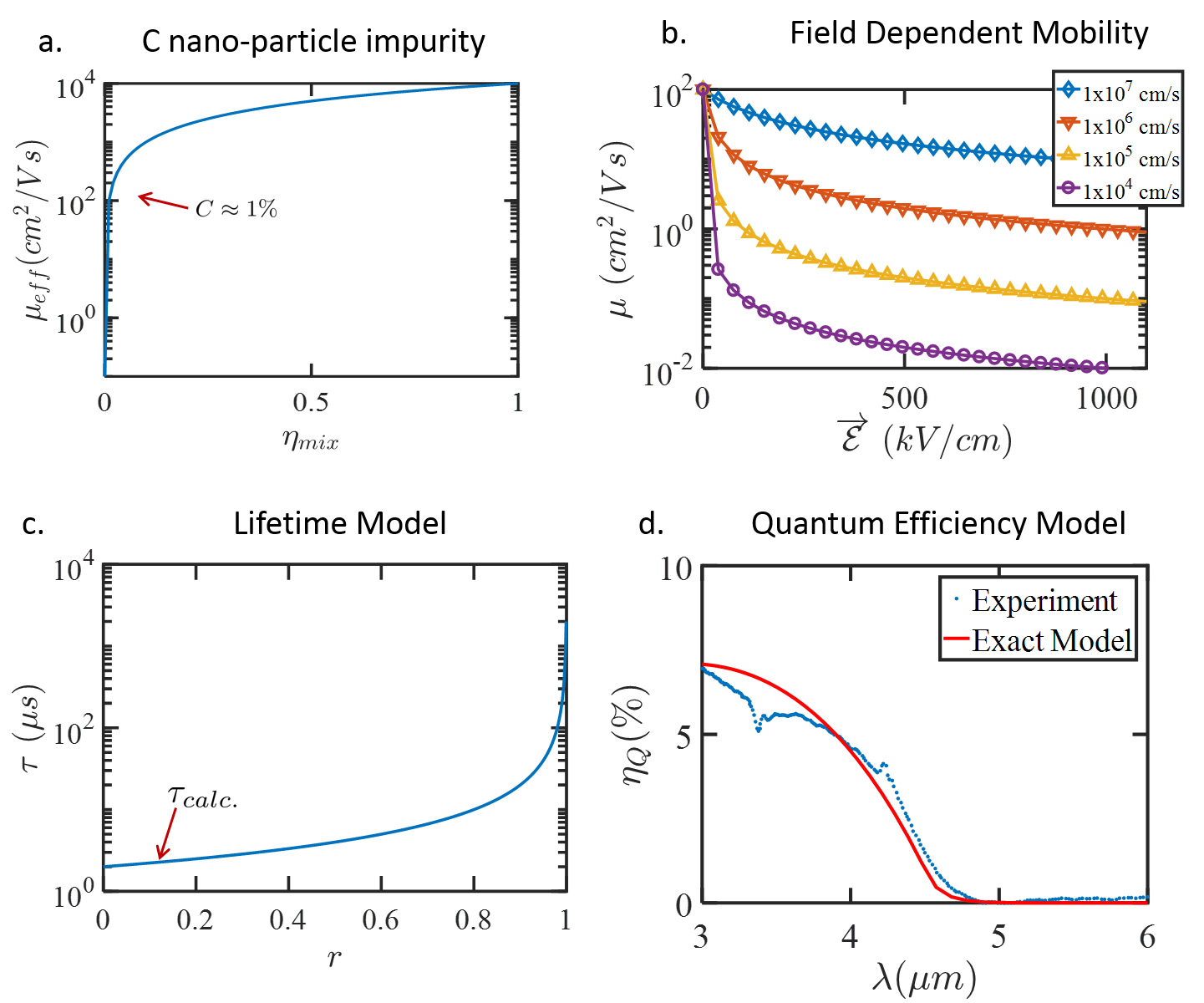}
\caption{b. Effect of $C$ nano-particles at the interfaces captured using a virtual crystal approximation. b. Field effect on mobility for various $V_{sat}$ using the Thomas-Caughey model.  c. Lifetime calculated for the effect of varying depletion widths. d. Benchmark of numerical or exact Quantum Efficiency (QE) model with experimentally measured QE. }
\label{fig:3}       
\end{figure}

From this model, it is easy to show (fig. \ref{fig:3}b) that even for a mix ratio $\chi=1\%$ it is possible to get the observed mobility. 

We also include the effect of applied electric-field on mobility using the well known Thomas-Caughey model for field dependent mobility \cite{klaassen_unified_1992} that has been used in a myriad of materials successfully and is given by:

\begin{equation}
\mu = \frac{\mu_0}{[1+(|\mu_0 \overrightarrow{\mathcal{E}}|/v_{sat})^\beta]^{1/\beta}}
\label{eq:mobility-fld}
\end{equation}

We have used $\beta = 1$ keeping with the literature on thin film polycrystalline materials (fig. \ref{fig:3}a). 

In the characterization data ($I_{dark}-V_{DS}$) available to us, we did not see a wide variation in mobility on the temperature in the range of interest from mid- to ambient-temperature operation, however empirical power law relationships have been developed elsewhere, see for \cite{schlichting_mobility_1973} an example. Additional dependencies of mobility such as carrier concentration and grain orientations ($111$ vs $100$) can be incorporated using power laws and virtual crystal models to obtain a universal mobility model for $PbSe$. 

\subsection{Lifetime}

Much like mobility, developing a universal lifetime model is a complex task especially in a polycrystalline film. The exact carrier lifetime details depend on the detailed charge distribution in the film, as well as various recombination phenomena, trapping and de-trapping of carriers in the inter-crystallite regions. In this work, we introduce one possible approach which can be expanded to incorporate these details numerically.

The total recombination rate given by the sum of the three principle recombination phenomena, radiative, Auger, Shockley-Reed Hall:

\begin{equation}
R_{tot} = r_{rad} np+r_{Aug;e}n^2p+r_{Aug;h}np^2 + \frac{N_{rec}}{\tau_{SRH;e}}+ \frac{P_{rec}}{\tau_{SRH;h}}
\label{eq:R-tot}
\end{equation}

Let us consider the effect of low level injection of carriers into the film and we want to calculate the lifetime through the change in recombination rate due to excess carriers. It can be shown that if we define a parameter $r_{surf} = V_{depletion}/V_{volume}$ which is the ratio of the average depleted crystallite volume (described previously in the bandstructure section), to the total crystallite volume, the lifetime can be written as :

\begin{equation}
\tau^{-1} = \frac{dR_{excess}}{d \Delta p} = \frac{\partial R}{\partial \Delta p} + \left(1-r_{surf}\right) \frac{\partial R}{\partial \Delta n}
\label{eq:liftime}
\end{equation}
 
where the individual partial derivatives are given by the base recombination rates:

\begin{eqnarray}
\frac{\partial R}{\partial \Delta p} &=& r_{rad}n_0 + r_{Aug;e} n_0^2 + 2r_{Aug;h}n_0p_0 \\
\frac{\partial R}{\partial \Delta n} &=& r_{rad}p_0 + 2 r_{Aug;e} n_0p_0 + r_{Aug;h}p_0^2
\label{eq:r-partials}
\end{eqnarray}

Fig. \ref{fig:3}c shows the effect of varying $r_{surf}$ on $\tau$. It can be seen that for estimated $r_{surf}(=0.1)$ of the films fabricated at NGC, the $\tau \approx 2\ \mu s$, which closely matches the extracted lifetime from $\mathcal{D}^*$ measurements, but is less than the lifetime directly measured from the film.

It should be noted that there is a large body of literature on modeling of recombination rates ($r_{rad}$, $r_{Aug;e}$, $r_{Aug;h}$ etc.) \cite{emtage_auger_1976,ziep_nonradiative_1980}, particularly on sensitized $PbSe$ which discusses the role of charge depletion in the crystallites on enhancing the lifetime of the majority carriers \cite{qiu_study_2013,zhao_understanding_2014}, which is reflected in high sensitivity of these films. These workers have also noted recently that the films have complex chemistry with a mix of varied species of $Pb, Se, O$ present in their films \cite{kumar_pbse_2017}, which hints back to classic explanation of high sensitivity by Humphrey \cite{humphrey_photoconductivity_1957} and Slater \cite{slater_barrier_1956} in terms of trapping of minority carriers in the inter-crystallite regions. Gaury and co-workers \cite{gaury_charged_2017} have found that for other polycrystalline photo-sensitive materials, charged grain boundaries can cause formation of potential wells where minority carrier trapping can happen, leading to reduced carrier recombination, and it is plausible that such a mechanism may play a role in $PbSe$ films as well.  

All these together make ab-initio modeling of carrier lifetime an enormously difficult task and our model attempts to bridge the above two theories together by postulating a transverse (to the film) transfer of electrons due to band-bending and depletion fields from the interface into the bulk of the crystallites, wherein they get trapped at the grain boundaries. We can expect a lot of sample-to-sample variability in the lifetime and its physics, depending on the growth conditions, as well as measurement approach, which may excite certain modes of recombination depending on the details of the characterization setup such as monochromatic high powered laser vs. weak blackbody incandescence \cite{gudaev_influence_1991}.

\begin{figure}[t]
\center
  \includegraphics[width=3.3in]{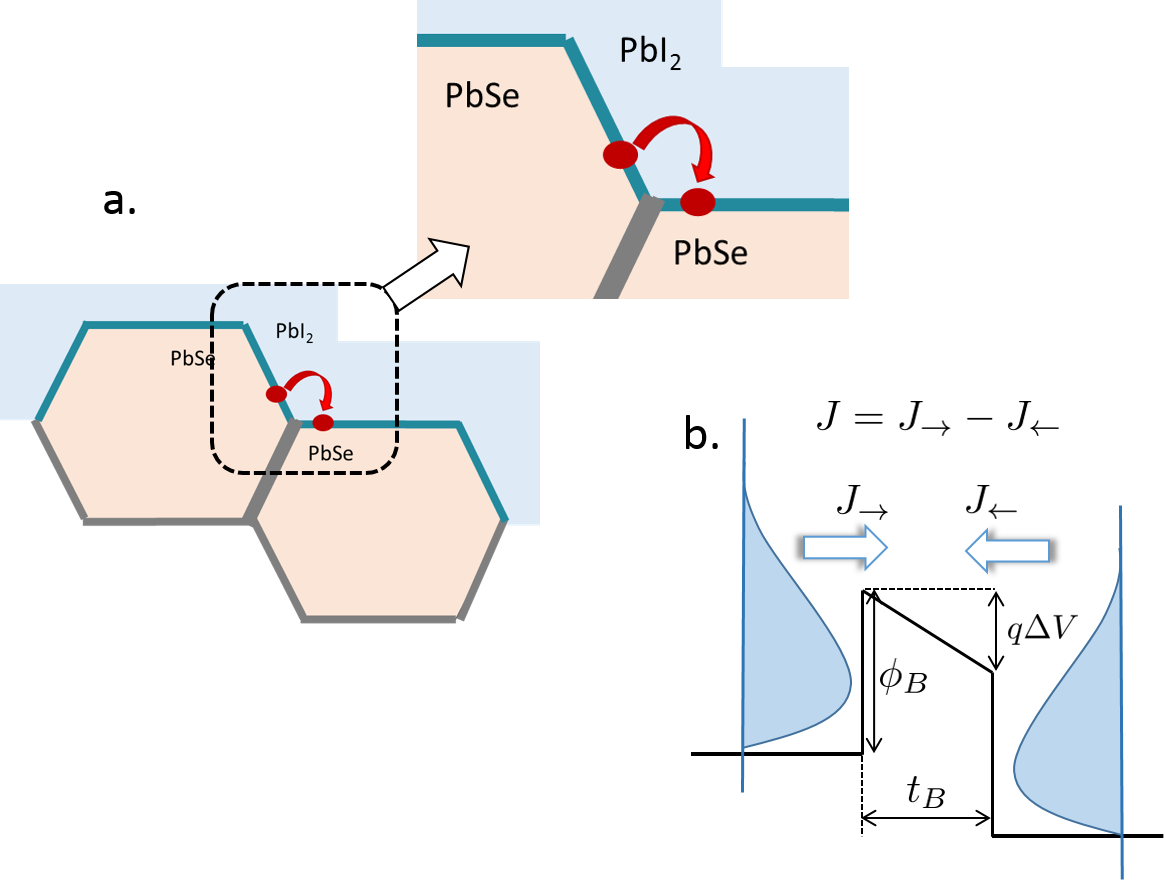}
\caption{a. Conduction mechanism of the interfacial holes (at $PbSe|PbI_2$) is through hopping over inter-crystallite barriers by thermionic emission. b. Landauer model of transport, where the net current density is given by the difference of right and left moving carriers over the barrier.}
\label{fig:4}       
\end{figure}

\section{Detector Transport Model}
\label{sec:transport}
\subsection{Dark Current}

In the physical picture developed so far, we have proposed that a transverse or ``vertical'' current separates the electrons away from the interfacial holes which reside in the potential well of the inverted channel, causing an increase in lifetime through suppression of Auger recombination mechanism. As discussed before, the conducting carriers, i.e. holes are confined to the interfacial inverted potential wells, at an individual crystallite level. Taking a step back and looking at the full detector film, this inverted surface will look like islands of inverted regions separated from each other by grain boundaries, which form a barrier for the holes. Therefore, we need to develop a model for the source-to-drain or the ``horizontal'' current in the film which flows along the $PbSe|PbI_2$ interface and is observed directly as the I-V characteristics of the detector. To do that, we model the current flow across a grain boundary barrier (GBB) as shown in fig.~\ref{fig:4}a-b. If the total applied voltage bias on the film is $V_{DS}$ and the average number of grain boundaries in the direction of the resulting current flow is $N_b$, the voltage drop $\Delta V$ across the GBB is given by:

\begin{equation}
\Delta V = \frac{V_{DS}}{N_b}
\label{eq:vdrop}
\end{equation}

and the resultant steady state net current density across the GBB can be expressed as the difference between the right flowing and the left flowing carriers given by:

\begin{equation}
\overrightarrow{J} = \overrightarrow{J}_{\rightarrow} - \overrightarrow{J}_{\leftarrow}
\label{eq:Jlr}
\end{equation} 

which can be related to the carrier densities on the left and the right grain/crystallites:

\begin{equation}
\overrightarrow{J} = q(p_l -p_r) \overrightarrow{v}
\label{eq:Jplr}
\end{equation}

The carriers concentration above the barriers at zero bias is given by (for the doping level $p_0$):

\begin{equation}
p = p_0\exp(\frac{-q\phi_b}{kT})
\label{eq:p_thermionic}
\end{equation} 

For the applied bias of $\Delta V$ across the barrier, we can write down the concentrations on left and right sides as 

\begin{equation}
p_{l;r} = p_0\exp(\frac{-q\phi_b\pm q\Delta V/2}{kT})
\label{eq:plr}
\end{equation}

Plugging this in eq.~\ref{eq:Jplr}:

\begin{eqnarray}
\overrightarrow{J} &=& qp_0[\exp(\frac{-q\phi_b+q\Delta V/2}{kT}) - \exp(\frac{-q\phi_b-q\Delta V/2}{kT}) ]\overrightarrow{v}\\
&=& 2qp_0\exp(\frac{-q\phi_b}{kT})\sinh(\frac{q\Delta V}{2kT})\mu\overrightarrow{\mathcal{E}}
\label{eq:currentdensity}
\end{eqnarray}

which can be further simplified as:

\begin{equation}
|J| = 2qp_0\exp(\frac{-q\phi_b}{kT})\sinh(\frac{qV_{DS}}{2N_bkT})\mu\frac{\phi_b}{t_b}
\label{eq:currentdensity-scalar}
\end{equation}
It is clear from the above expression that the density of the carriers are modulated by the barrier height and the applied voltage difference, while the velocity is dictated by the barrier properties (mobility and the electric field across it). 

The barrier height $\phi_b$ depends on the chemistry of the GBBs. In other polycrystalline photo-sensitive materials ($CdTe, Cu(In,Ga)Se_2$) they play an important role in controlling the dark current in photo-voltaic devices (solar cells) \cite{gaury_charged_2016}. Even though $PbSe$ is operated in photoconductive mode, the $\phi_b$ is still a critical parameter in the $I-V$ characteristics, for both dark and photo-generated current. For model validation purposes, we have used it as a fitting parameter, and further structural investigations and electrical characterization can help in understanding the detailed chemistry of GBBs and their implications for $PbSe$ based photo detectors.

The model presented above is based on the Landauer picture of conduction \cite{datta_electronic_1997}, where we have assumed that the voltage drop is across the GBBs only, while the grains themselves act as ``reservoirs'', and the whole detector itself is a network of small Landauer conductors \cite{datta_lessons_2012,ghosh_nanoelectronics:_2016}. Therefore, variability effects can be numerically captured using Monte-Carlo simulations of Landauer conductor networks.

In this model we have assumed that isolated $PbSe$ crystallites embedded in the $PbI_2$ layer do not conduct due to large barrier presented by the surrounding $PbI_2$ film. There is still a possibility of a small amount of electrons to tunnel through and add to the total current flow. However, we have ignored that contribution at present since in the benchmarks performed later in the paper we seem to be able to capture the $I-V$ characteristics of test detectors adequately without this tunneling current contribution, which possibly is vanishingly small in well made detectors we have worked with.

\subsection{Photo-Generated Current}

From the eq.~\ref{eq:currentdensity-scalar}, it is clear that the primary determining factor behind the conduction in the material is the modulation of the charge carriers. The optical irradiation causes generation of extra carriers which increase the conductivity of the film. If the quantum efficiency is given by $\eta(\lambda,T,x)$ (dependencies on wavelength $\lambda$, temperature $T$, and the depth in the film $x$), the total rate of generation of carriers in the film is given by:

\begin{equation}
G_l(T) = \int_0^t\int_0^\infty{\eta(\lambda,T,x)\frac{P_{opt}(\lambda)}{\frac{hc}{\lambda}x}}d\lambda dx
\label{eq:gl}
\end{equation}

which can be numerically integrated. For low level generation, the total number of generated carriers per unit volume is given by: 
\begin{equation}
\Delta p \approx G_l\tau_p/t_{film}
\label{eq:deltap}
\end{equation}

This extra generation of carriers due to the optical gating can be considered as a channel enhancement, and larger the lifetime is higher the enhancement. The illuminated/light current density is then given by:

\begin{equation}
|J| = 2q(p_0+G_l\tau_p/t_{film})\exp(\frac{-q\phi_b}{kT})\sinh(\frac{qV_{DS}}{2N_bkT})\mu\frac{\phi_b}{t_b}
\label{eq:lightcurrentdensity-scalar}
\end{equation}

It is illustrative to compare the transport model developed here with that of solar cells and photovoltaic detectors. Even though all these devices depend of photon incident to modulate the current in them, the dominant transport mechanism in solar cells and photovoltaic detectors is bipolar minority carrier diffusion where the mode of operation is decided by the applied bias on the diode like device (forward bias for solar cells, reverse bias for detector), whereas the transport mechanism of photoconductors is majority carrier drift (by a hopping mechanism) which makes it work like a Field Effect Transistor (FET) rather than a diode. Therefore, the I-V characteristics of photoconductives are very different than the photovoltaics. The I-V of photoconductors pass through the $(V,I) = (0,0)$ point with different slopes for dark and photo-generated currents, whereas the photo-generated current of photovoltaics do not pass through $(V,I) = (0,0)$ but rather through the $(V,I) = (I_{sc},0)$, where $I_{sc}$ is the short circuit current caused by chemical gradients ($\nabla \mu_{e,h}$) rather than electrostatic gradient ($\nabla \phi$).

It is also illustrative to compare $PbSe$ photoconductors with inverted channel Field Effect Transistor (FET). In this comparison, the role of the threshold gate voltage in FET is played by the degree of band bending at the interface of the detector, any extra gate voltage is analogous to the optical power incident on the detector, the electrostatic control of the gate  over the inverted channel (characterized by gate transconductance) is equivalent to the quantum efficiency of photo-carrier generation.

We have put these comparisons together in table \ref{tab:comparison} for easy review.

\begin{table*}[ht]
\caption{Comparative description of $PbSe$ detector physics}
\label{tab:comparison}
\centering
\begin{tabular}{|c|c|c|}
\hline
\multicolumn{3}{|c|}{Photoconductive vs. Photovoltaic} \\\hline\hline
 & Photoconductive & Photovoltaic \\\hline
Carrier type & unipolar & bipolar\\
Transport gradients & electrostatic ($\nabla \phi$) & chemical  ($\nabla \mu_{e;h}$) \\
Transport type & majority carrier hopping (drift) & minority carrier diffusion \\
$I_{photo}\ vs\ V$ & through $(V,I)=(0,0)$ & through $(V,I)=(0,I_{sc})$ or $(V,I)=(V_{oc},0)$\\
Detection method & change in $R$ & change in $I_{sc}$\\
Device analogy & field effect transistor (FET) & diode\\\hline\hline
\multicolumn{3}{|c|}{$PbSe$ photoconductor vs. MOSFET} \\\hline\hline
 & $PbSe$ photoconductor & MOSFET \\\hline
Doping & electrostatic & chemical\\
Carrier inversion & intrinsic band-bending & extrinsic gate voltage\\
Channel Resistance change & photo-carrier generation & channel surface potential \\
Current modulation & quantum efficiency & gate transconductance\\\hline
\end{tabular}
\end{table*}

\subsection{Quantum Efficiency}

The goal of the QE model is the capture the photon-charge carrier interaction, in particular the carrier generation rate. In our approach, we have used an inelastic photo-generation model, inspired by the Shockley-Queisser approach \cite{shockley_detailed_1961}, where the photons of energy higher than band gap ($E>E_g$) thermalize to the band edge and produce the photo-generated carriers. In this case the QE is numerically given as:

\begin{equation}
    \eta(\lambda,T_{det}) = \frac{E_g\int D(E)f(E)dE}{\int E D(E)f(E)dE}
\end{equation}

Here $D(E)$ and $f(E)$ are the photon density of states and occupancy (given by Bose-Einstein statistics) functions.

We have also considered the effect the disorder in the polycrystalline material, captured through the Urbach tails into the band gap \cite{urbach_long-wavelength_1953}, which allows sub-$Eg$ photons to generate carriers as well. The absorption is given as a decaying exponential, with a fitted Urbach energy $E_{Urb}$:

\begin{equation}
    \alpha \propto \exp(\frac{E-E_g}{E_{Urb}})
\end{equation}

We account for the non-uniformity of absorption in the film thickness direction ($x$) using the classic Moss model \cite{moss_photoconductivity_1965} given by:

\begin{equation}
 \eta(x) = \frac{(1-r)(1-e^{\alpha x})}{1-re^{\alpha x}}
\end{equation}

where $r$ is the reflectivity on the sample surface, while $\alpha$ is the effective absorption coefficient.

While these three together produce the IQE (internal QE), the EQE (external QE) is typically less than the IQE, due to reflection from the surface, and can vary depending on the anti-reflection coating (ARC) quality, which can be accounted for separately.

The above ``exact model'' needs to be solved numerically to calculate the total generation rate (eq.~\ref{eq:gl}) and the light current (eq.~\ref{eq:lightcurrentdensity-scalar}). We have developed a phenomenological model inspired from high-order Chebyshev filter transfer functions and is given by:

\begin{equation}
    \eta = \frac{\eta_0}{1+\gamma\lambda^{k}}
\end{equation}

where both the parameters $\eta_0$ and $\gamma$ collectively capture the effect of temperature and film thickness. As a benchmark, we have compared experimentally measured QE with the exact model in fig. \ref{fig:3}d with a close match.

We have summarized the parameters of the model in table \ref{tab:material}.

\begin{table*}[ht]
\caption{Parameters of the model}
\label{tab:material}
\centering
\begin{tabular}{|c|c|c|}
\hline
Parameter & Symbol & Dependencies \\
\hline\hline
\multicolumn{3}{|c|}{Material Parameters} \\\hline
Bandgap & $E_g$ & material,$\ T$\\
Carrier Concentration & $p_0$ & process \\
Average Grain and Intergrain lengths & $l_g,l_b$ & process \\
Average Intergrain Barrier height & $\phi_b$ & process\\
Workfunction & $\Phi_s$ & $\chi_s, p_0,E_g, T$\\\hline
\multicolumn{3}{|c|}{Transport Parameters} \\\hline
Mobility & $\mu$ & $\mathcal{E}, T,p_0,\xi_{mix}$\\
Quantum Efficiency &$\eta$& $\lambda,T,d$\\
Lifetime &$\tau_p$& $r_{surf},T,N_T,E_T$\\\hline
\multicolumn{3}{|c|}{Detector Parameters} \\\hline
Contact resistance & $R_c$ & $\rho_c,\rho_M,p_0,\phi_M,\chi_s, T$, detector design  \\
Noise & - & $T,p_0$, detector design\\
\hline
\end{tabular}
\end{table*}

\section{Full Detector Model}
\label{sec:detector}

To fully capture the electrical behavior of the detectors and their performance when used in a focal plane array (FPA), it is necessary to include the effect of the contact resistance and the noise in them. We briefly describe our approach to modeling these components.

\subsection{Contacts} 

The primary role of contacts is to add extra parasitic resistance to the current flow in the detector. Since the contact resistance is light insensitive, it reduces the overall photo-sensitivity of the detector looking from the electrical readout (ROIC). The aim of contact modeling is to capture this extra parasitic resistance which depends on the contact materials being chosen, effective doping in the $PbSe$ film, as well as operating conditions such as temperature and biasing voltage.

\begin{figure}[t]
\center
  \includegraphics[width=3.3in]{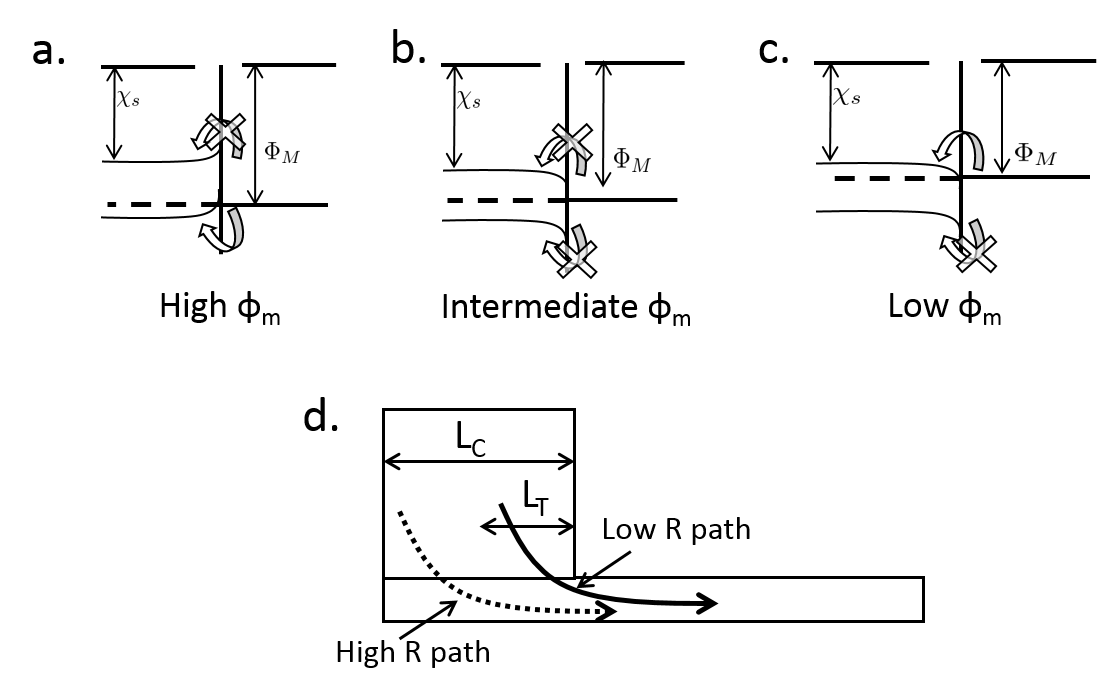}
\caption{a. High workfunction metal contact with a semiconductor forms a natural hole injector/absorber and a blocking contact for electrons due to Schottky barrier. b. Intermediate workfunction metal contact with a semiconductor forms a Schottky barrier with both electrons and holes. c. Low workfunction metal contact with a semiconductor forms a blocking contact for holes but Ohmic for electrons. d. The Effective Transfer length ($L_T$) or Injection/Absorption length from contact into the semiconductor is less than the physical contact length ($L_C$) due to current crowding at the edge caused by high resistance path at the outer edge of the interface.}
\label{fig:5}       
\end{figure}

The contact itself may be formed either through direct deposition on the $PbSe$ after removing the insulating $PbI_2$ layer or through a bottom metal oxide layer, possibly formed due to oxidation during $PbSe$ CBD and sensitization. The details of the contact model would vary depending on the particular contact development process and the chemistry. In general contact modeling is a complex and open modeling problem with a rich literature, therefore we only briefly discuss the case of metal contacts directly deposited on the $PbSe$ films.

Depending on the material used to form the contacts, the electrical nature of the contact can be either Ohmic or be a Schottky type. Using a bandstructure model of the contact, different metals form different types of contacts as shown in fig.~\ref{fig:5}a-c. It can be seen that metals such as $Ni, Pd, Mo, W$ etc. with large workfunctions are expected to naturally form Ohmic or very low barrier contacts with \textit{p}-type semiconductors due to band alignment, assuming ideal metallurgical contacts that can be described well using bulk workfunctions. These large workfunction metals also form high barrier or ``blocking'' contacts for electrons, which is useful in preventing sweep-out under high bias and contributes to larger photoconductive gain.

We use Richardson's equation to calculate the areal resistivity of the $PbSe|M$ interface whose workfunction difference is $\Delta\Phi$:

\begin{equation}
\rho_{int} = \frac{k}{qTA^*}\exp(\frac{\Delta\Phi}{kT})
\label{eq:rho-contact}
\end{equation}

Another important aspect of the contact resistance is the current crowding observed in resistive channels (fig.~\ref{fig:5}d). For highly resistive films, the current injection is not uniform under the contact pad, and the effective injection area is characterized by an transfer length $L_T$ for the charge which is smaller than the physical length $L_{lead}$ and is given by \cite{kennedy_two-dimensional_1968}:

\begin{equation}
L_T = \sqrt{\rho_c/r_s}
\label{eq:LT-contact}
\end{equation}

Where $\rho_c$ is the specific contact resistivity, and $r_s$ is sheet resistance.

Therefore the total effective resistance of the contact is given by:
\begin{equation}
R_{c} = \rho_{int}W_{lead}L_T + \rho_Mt_{lead}/L_{lead}W_{lead}
\label{eq:r-contact}
\end{equation}

Where the $W_{lead}$ is the total width of source and the drain pads, $t_{lead}$, $L_{lead}$ are the physical contact pad thickness and lengths, and $\rho_M$ is the contact metal volume resistivity.

For test detectors fabricated at Northrop Grumman, we found that the ratio of contact resistance for electrons-to-holes was $R_c^e/R_c^h\approx 10^7$ due to the choice of metals used which formed naturally blocking contacts for electrons. This was additionally confirmed with a) TLM measurements which showed well-formed Ohmic contacts for majority/conducting carriers (holes), and b) $\mathcal{R}-\mathcal{E}$ measurements which did not show any saturation even for very high applied fields indicating the lack of minority carrier (electron) sweep-out due to these naturally blocking contacts.

\subsection{Noise} 

We model the total rms noise spectra in the system as the sum of various noise processes in the detector. 

\begin{equation}
S_{n} = \sum_k a_ks_{k}
\label{eq:tot-noise}
\end{equation}

Where the individual noise spectra are given by \cite{konczakowska_noise_2011}:

\begin{eqnarray}
s_{thermal} &=& \frac{4kT}{R}\\
s_{shot} &=& 2q|I|\\
s_{1/f} &=& \frac{1}{f}\frac{\alpha_HI^2}{p\Omega}\\
s_{G-R} &=& \frac{4I^2}{p(1+4\pi^2f^2\tau^2)}
\label{eq:individual-noise}
\end{eqnarray}

Using a phenomenological weighing of each of these individual noise types, obtained from noise measurements, it is possible to match the total noise and then use the noise model to predict SNR and the specific detectivity $\mathcal D^*$ of the detector. When the noise in the system is predominantly due to the background photon flux noise, the detector is said to be working in the BLIP limit which is theoretically the best possible performance obtainable from the detector.

The total noise current then can be calculated as:

\begin{equation}
I_{n} = \sqrt{S_n\Delta f}
\label{eq:tot-noise-cur}
\end{equation}

Where $\Delta f$ is the measurement spectral bandwidth.

The noise model is used to benchmark the $\mathcal{D}^*$ of the detectors which is discussed later in this paper.

\begin{figure}[t]
\center
  \includegraphics[width=3.3in]{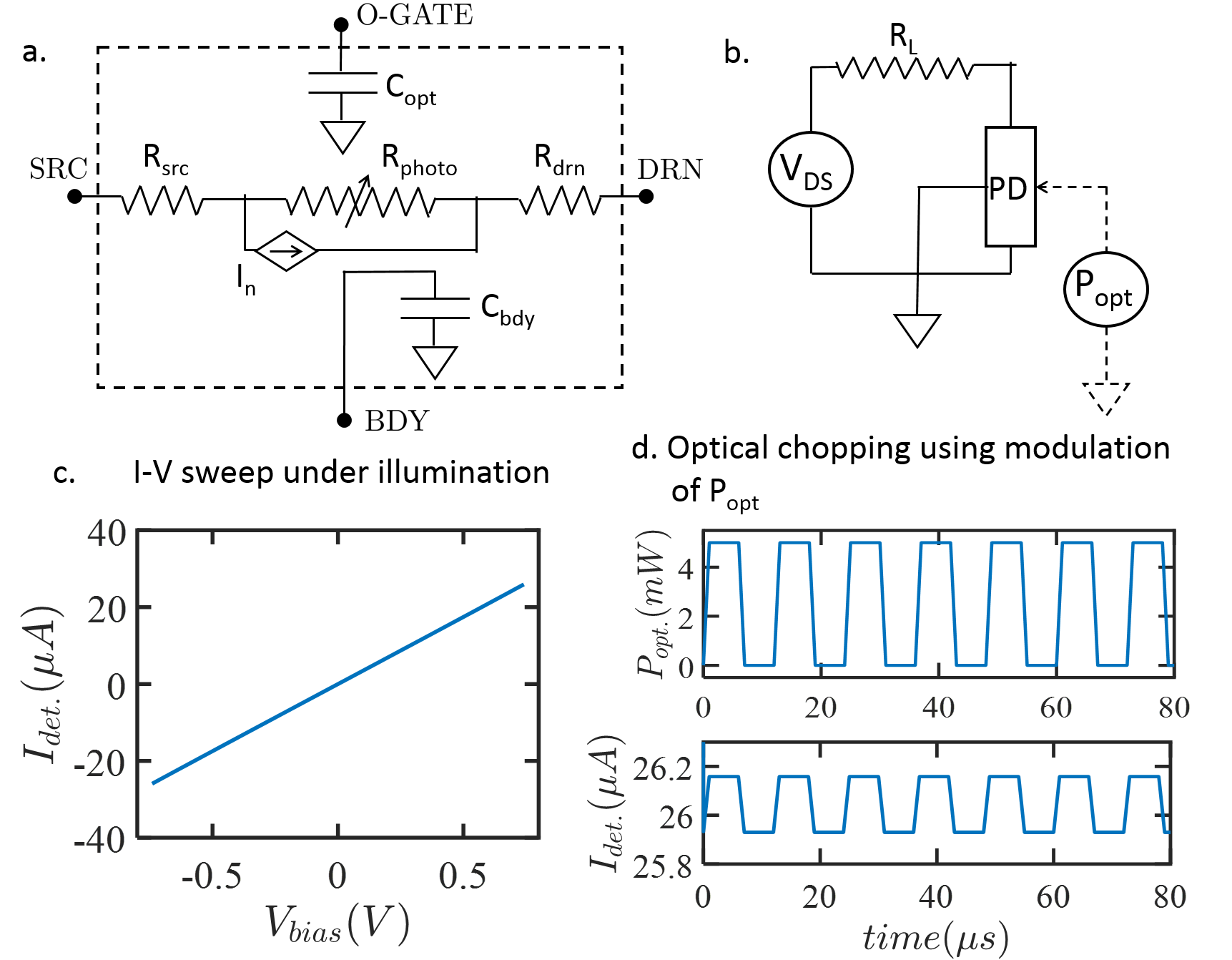}
\caption{a. Circuit model for the photo-detector b. Circuit-testbench used to perform simulations c. $I_{DS}-V_{DS}$ obtained by dc sweep. d. $I_{DS}-P_{opt}$ obtained by a transient simulation for a fixed $V_{DS}$ but modulated optical excitation on the $O-GATE$ terminal, mimicking the effect of an optical chopper.}
\label{fig:6}       
\end{figure}

\section{Detector Circuit Model}
\label{sec:circuit}

We develop a Verilog-A based SPICE compatible compact model that can be used to perform large scale FPA-ROIC simulations, for both analog and digital ROICs. The circuit model for the photodetector is shown in fig.~\ref{fig:6}a where the $SRC$ and $DRN$ terminals represent the physical contact pads. The $O-GATE$ terminal represents the optical ``terminal'' to which a voltage source may be applied to provide the optical power incident on the detector. The extra $BDY$ terminal is provided to account for a body bias/back gate voltage that may be provided from a ROIC to control the electrostatics of the detector. 

The photodetector is implemented as a controllable resistor with a parallel controllable current source. The optical incident power and the back gate voltage control the resistor, while the current source is tuned to produce the correct amount of noise current in the detector. In addition we have also incorporated two extra resistors to account for the contact resistances (source and drain).

SPICE compatible compact models open up a wide variety of circuit simulation and detector-readout codesign capabilities that may not be easily available in other software, by leveraging the capabilities developed in circuit simulators over many decades of development. As an example, using the same circuit testbench (fig.~\ref{fig:6}b) we can obtain a DC bias simulation (fig.~\ref{fig:6}c), including noise, as well as a transient sweep of optical power modulation (through a chopper) on the detector current (fig.~\ref{fig:6}d). Other examples of simulations available are Monte-Carlo based simulations for variability analyses, ac and transient noise analysis, as well as measure based dissipation and self-heating analyses.

\section{Model Verification and Performance Projection}
\label{sec:projection}

\begin{figure}[t]
\center
  \includegraphics[width=3.3in]{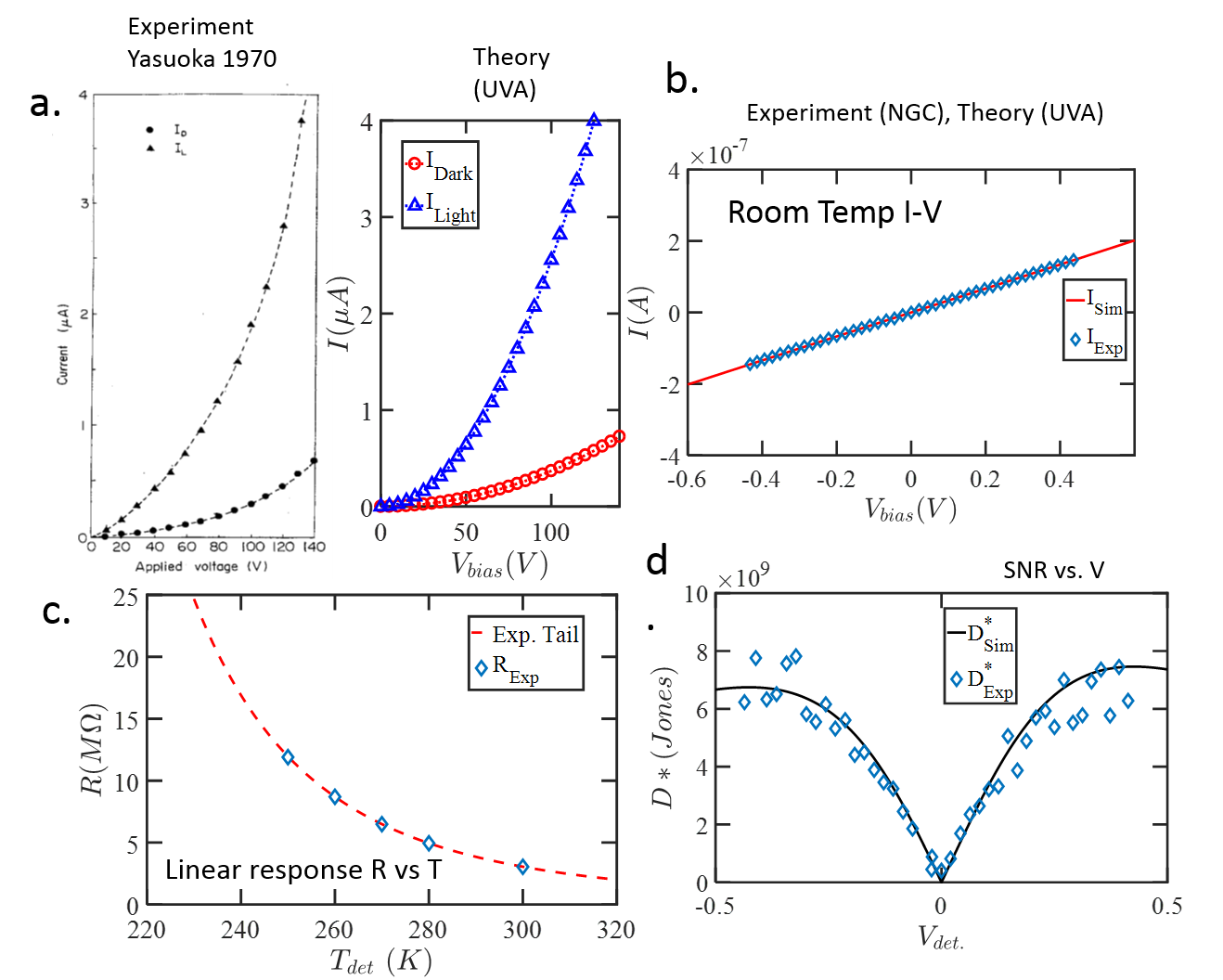}
\caption{a. Benchmarking dark and light $I-V$ of a classic experiment (1970) b. Benchmarking room temperature $I-V$ of detectors fabricated at Northrop Grumman Corp. (2017) c. Linear response $R-T$ benchmarked with NGC detectors d. Benchmarking of $\mathcal {D}^* - V$ between theoretical model and the experimentally measured detector performance.}
\label{fig:7}       
\end{figure}

We use two different and independent set of experimental data on $PbSe$ photodetectors to benchmark our model, one being a classic experiment available in literature \cite{yasuoka_photoconductivity_1970,yasuoka_effects_1968}, and the other being $18 \mu m \times 100\mu m$ sized detectors fabricated at Northrop Grumman \cite{yoo_high-operating-temperature_2017}. We use material parameters and film geometry directly measured, inferred from reported data, or using values close to those reported for other similar experiments on $PbSe$ in literature, as required. As shown in fig. \ref{fig:7} we obtain very close match to $I-V$ characteristics for both low as well as high bias regimes of the two different experiments, using the same model we have developed, but using parameters for their respective experiments. In addition we also closely match $R-T$ characteristics, and $\mathcal D^*-V$ trends quantitatively. This demonstrates the validity of the modeling approach, the physicality of the model developed, as well as its utility as a benchmarking and exploratory platform to investigate performance enhancement through material improvements and selecting optimal operating conditions.

\begin{figure}[t]
\center
  \includegraphics[width=3.3in]{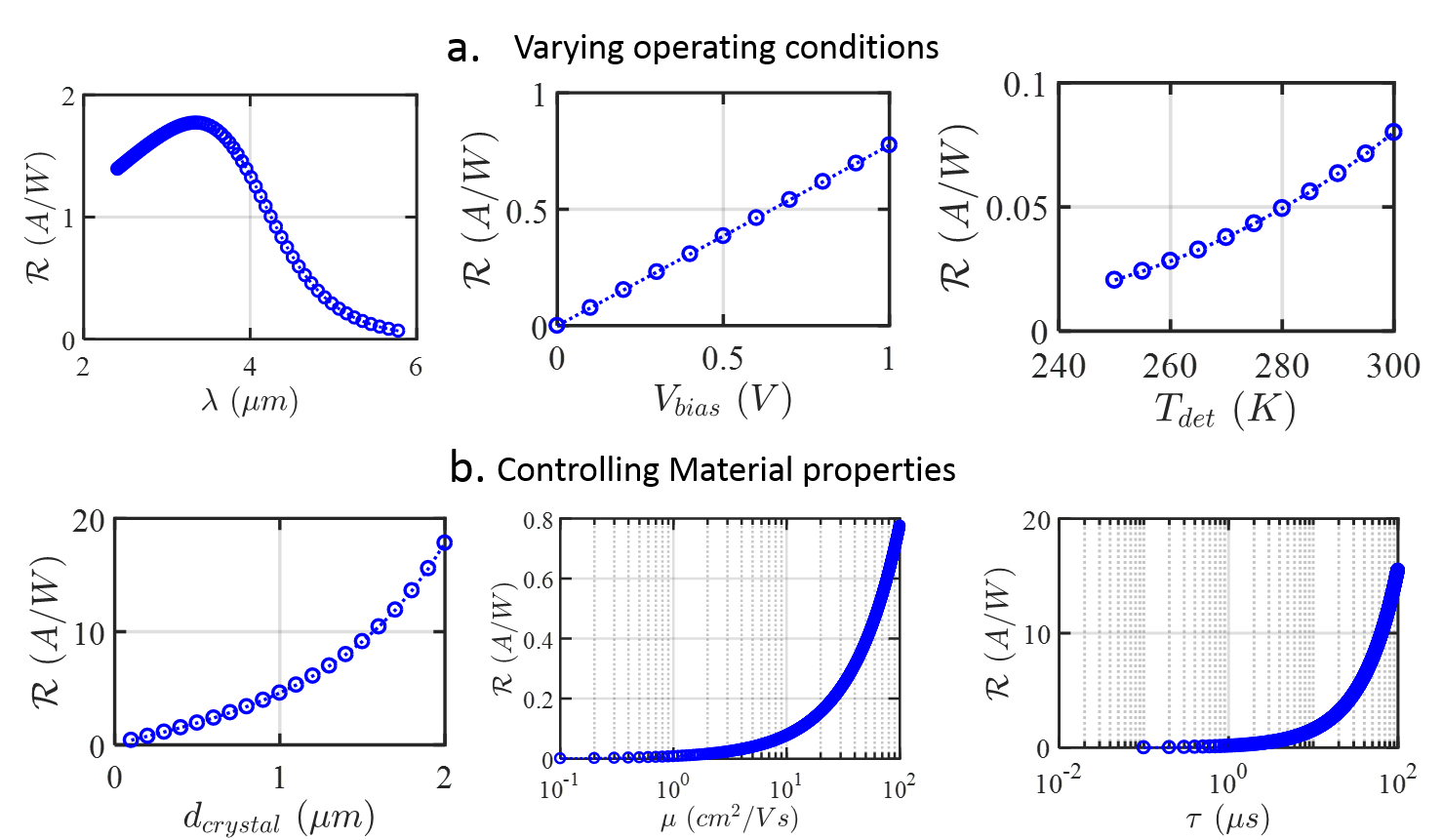}
\caption{a. Exploring variation in operating conditions (wavelength $\lambda$, bias $V_{bias}$, detector temperature $T_{det}$) on responsivity $\mathcal R$ b. Exploring variation in material properties (average crystallite size $d_{crystal}$, carrier mobility $\mu$, carrier lifetime $\tau$) on responsivity $\mathcal R$. }
\label{fig:8}       
\end{figure}

As an example of possible performance analysis, we look at the responsivity $\mathcal R$ (defined as $I/P_{opt.}$) of detectors whose parameters are close to (but not exactly the same) those of the Northrop Grumman test detectors. We first vary a few operating parameters such as wavelength (keeping within the MWIR range), ROIC bias, and detector dewar temperature, and plot the change in the responsivity. We show these in fig.~\ref{fig:8}a which captures the projection as obtained from the model. We also test the effect of the change of material properties that may be obtained from varying fabrication recipes and conditions. We look at the effect of variation in crystallite size, mobility, and majority carrier lifetime and plot these results in fig.~\ref{fig:8}b. 

The $\mathcal{R}-\lambda$ shows the responsivity goes through a peak close to but slightly shifted from the band edge, closely following the QE mechanism discussed before. It also shows the effect of Urbach tail as a broadening above the cutoff $\lambda$. The $\mathcal{R}-T$ curve shows an increase due to thermal broadening of the Fermi (carrier) distribution above the GBBs, while the $\mathcal{R}-V$ captures the effect of increased electric field due to increased bias voltage. It should be noted that higher $\mathcal{R}$ at higher $T$ does not necessarily mean better photodetector performance, since higher $T$ also concomitantly causes higher dark current, as well as increased noise in the detector. The $\mathcal{R}-d_{crystal}$ captures the effect of increasing field across GBBs as the number of voltage drop points are decreased for the same bias, $\mathcal{R}-\mu$ captures the effect of increased mobility resulting in higher photo-generated current, while $\mathcal{R}-\tau$ captures the higher responsivity due to higher photogain.

This exercise shows the complex phase space of parameters that exist for the photodetectors that can be accurately captured by our modeling approach, connecting fundamental material properties and disorders to circuit level performance, seamlessly.

\section{Looking to the Future}
\label{sec:conclusion}

In this paper, we have presented a multi-scale physics based framework for materials-to-systems modeling of photodetectors. We used an example of $PbSe$ based MWIR photodetectors which were fabricated and characterized by our experimental team at Northrop Grumman and University of Virginia. We built a set of interconnected models all the way from the bandstructure of the sensitized film to the SPICE based compact model that can capture the effect of material variability. We further validated the model on two independent experiments, as well as different physical characteristics of the detector and illustrated the exploratory nature of the approach by analyzing the effect of varying material properties and operating conditions of system level performance parameter.

Such an approach fits closely with the fast emerging trend of monolitihic integration of ultra-scaled photodetectors with DSP ROICs leveraging the advances in CMOS for close-to-sensor data processing. The IoT market is driving the demand of integrating neuromorphic processing and deep-learning into the world of sensors. These methods have been demonstrated to be able to make inference from video data and can be implemented on an FPGA board integrated with a detector-ROIC monolithic chip. We can expect to see more sophisticated ``smart camera'' backends that will be integrated with advanced photonic devices in near future. See \cite{ganguly_hardware_2018} as an example.

We believe that the materials-to-systems modeling approach presented by us fits into these trends and enables an integrated platform for co-design of electronics and photonics. 

\section*{Acknowledgments}
The authors would like to thank Mr. Justin Grayer for invaluable discussions. This research was sponsored by DARPA/MTO through the WIRED program Contract No. FA8650-16-C-7637

\bibliographystyle{ieeetran}
\bibliography{PbSe}

\end{document}